\documentclass[aps,prc,twocolumn,showpacs,preprintnumbers]{revtex4}

\usepackage{graphicx}
\usepackage{dcolumn}
\usepackage{bm}

\usepackage{amsmath}
\usepackage{amstext}
\usepackage{amssymb}
\usepackage{natbib}

\begin{document}

\title{$^{1}S_{0}$ proton superfluidity in neutron star matter:
  Impact of bulk properties}

\author{Tomonori Tanigawa}
\altaffiliation[Mailing address:~]{%
  Japan Atomic Energy Research Institute, Tokai, Ibaraki 319-1195,
  Japan%
}%
\email[Email address:~]{tanigawa@tiger02.tokai.jaeri.go.jp}
\affiliation{%
  Japan Society for the Promotion of Science, Chiyoda-ku, Tokyo
  102-8471, Japan%
}%
\affiliation{%
  Advanced Science Research Center, Japan Atomic Energy Research
  Institute, Tokai, Ibaraki 319-1195, Japan%
}%
\author{Masayuki Matsuzaki}
\email[Email address:~]{matsuza@fukuoka-edu.ac.jp}
\affiliation{%
  Department of Physics, Fukuoka University of Education, Munakata,
  Fukuoka 811-4192, Japan%
}%
\author{Satoshi Chiba}
\email[Email address:~]{sachiba@popsvr.tokai.jaeri.go.jp}
\affiliation{%
  Advanced Science Research Center, Japan Atomic Energy Research
  Institute, Tokai, Ibaraki 319-1195, Japan%
}%

\date{\today}

\begin{abstract}
  We study the $^{1}S_{0}$ proton pairing gap in neutron star matter
  putting emphasis on influence of the Dirac effective mass and the
  proton fraction on the gap within the relativistic
  Hartree-Bogoliubov model. The gap equation is solved using the
  Bonn-\textit{B} potential as a particle-particle channel interaction.
  It is found that the maximal pairing gap $\Delta_\mathrm{max}$ is
  1--2~MeV, which has a strong correlation with the Dirac effective
  mass. Hence we suggest that it serves as a guide to narrow down
  parameter sets of the relativistic effective field theory.
  Furthermore, the more slowly protons increase with density in the
  core region of neutron stars, the wider the superfluid range and the
  slightly lower the peak of the gap become.
\end{abstract}


\maketitle

\section{Introduction}
\label{sec:introduction}

Superfluidity in neutron star matter is one of the hot issues in
nuclear astrophysics since superfluidity plays a key role in affecting
the cooling of neutron stars~%
\cite{kunihiro93:_various_phases_high_nuclear_matter_neutr_stars}.
Unlike the $^{1}S_{0}$ neutron pairing, the $^{1}S_{0}$ proton pairing
occurs in dense matter with supranuclear density, where its
properties are not much known. Such a region is
highly relevant to the URCA processes that control cooling of neutron
stars. 

In neutron stars, several types of baryon pairing are 
believed to appear. In the inner crust region, neutrons will form
the $^{1}S_{0}$
pairs~\cite{takatsuka93:_super,wambach93:_quasip,chen93:_pairin}. At
the corresponding baryon density $10^{-3}\rho_{0} \lesssim \rho_{B}
\lesssim 0.7\rho_{0}$, where $\rho_{0} \simeq 0.15$~fm$^{-3}$ is the
saturation density of symmetric
nuclear matter, the $^{1}S_{0}$ partial wave of the nucleon-nucleon
(\textit{NN}) interaction is attractive. This attraction helps
neutrons pair up through the well-known BCS mechanism.  In the core
region $\rho_{B} \gtrsim 0.7\rho_{0}$, the $^{3}P_{2}$ neutron
pairing may also appear since the $^{3}P_{2}$ partial wave of the
\textit{NN} interaction becomes attractive
enough~\cite{takatsuka93:_super,elgaroey96:_tripl}. In contrast, the
$^{1}S_{0}$ partial wave would become repulsive there so that the
neutrons would cease to pair in the $^{1}S_{0}$ state. Instead, the
$^{1}S_{0}$ proton pairs are predicted to appear owing to its fraction
smaller than neutrons~\cite{takatsuka93:_super,chen93:_pairin}.
At much more higher baryon density $\rho_{B} \gtrsim 2\rho_{0}$,
various hyperons may emerge; some kinds of them possibly form
pairs in the same way as nucleons do~\cite{balberg98:_s_lambd,%
takatsuka99:_super_lambd_hyper_admix_neutr_star_cores,%
tanigawa03:_possib_lambd_hartr_bogol}. We, however, stay on a picture of
neutron stars without hyperons in the present study.

The size of the nuclear pairing gap in dense matter is controversial
since many uncertainties remain regarding \textit{NN} interactions in
dense medium, methods of approximation, sparsity of experimental data
under extreme conditions, and so on. From the perspective of the
present status of immaturity, a great deal of study with diversity is
absolutely necessary.
In the meantime, many studies of neutron stars have been performed
using various frameworks so far. Recently, relativistic models are
attracting attention in researches on neutron stars since they are
suitable to describe the stars in compliance with the special
relativity~\cite{glendenning00:_compac_stars}.
Most often used among them is the relativistic mean field (RMF) model,
particularly owing to its economical way of description. Hence we
choose the extended model of it, the relativistic Hartree-Bogoliubov
(RHB) model as the framework of the present study.

The primary aim of this paper is to elucidate effects of bulk
properties on the $^{1}S_{0}$ proton pairing correlation in neutron
star matter (consisting of \textit{n}, \textit{p}, \textit{e}$^{-}$,
and $\mu^{-}$) using the RHB model with capability of handling the
pairing correlation. Here we show importance of environmental
properties of dense matter that surrounds Cooper pairs of protons (the
pairs themselves are of course a part of the environment). In
particular, the Dirac effective mass of nucleons and the proton
fraction are considered as accompanying quantities of great importance
since the bulk properties are influential on superfluidity in neutron
stars. Therefore we aim to address the bulk properties of neutron star
matter and its superfluidity on the same footing within the RHB model.

The present study covers the two aforementioned quantities:
The Dirac effective mass has an effect on the pairing correlation via
the density of states.
The symmetry energy coefficient controls the proton fraction in neutron
star matter, especially its density dependence.
Needless to say, they connect with the equation of state (EOS) of
dense matter; it is strongly related to properties of
neutron stars, such as their internal structure, mass, radius, and
so on. These macroscopic properties of neutron stars set stringent microscopic
requirements for constituents; hence the pairing correlation is no exception.
In this study, we use two distinct RMF models with two distinct
parameter sets for each to compose neutron star matter; thereby we
have four distinct EOSs here. Two of them are for a comparison of the
effect of the Dirac effective mass, the other two are for a comparison
of the effect of the proton fraction.

This paper is organized as follows. In Sec.~\ref{sec:models}, model
Lagrangians and the gap equation for the $^{1}S_{0}$ proton
superfluidity are illustrated. In Sec.~\ref{sec:results-discussion},
results of the pairing properties in neutron star matter are
presented. Section~\ref{sec:summary} contains our summary and
conclusions.

\section{Models}
\label{sec:models}


\begin{table*}[tbp]
  \centering
  \caption{Parameter sets used in the present study. We fix the
    bare nucleon mass $M=939.0$~MeV and the $\rho$ meson mass
    $m_{\rho}=763.0$~MeV, except for TM1 where $M=938.0$~MeV and
    $m_{\rho}=770.0$~MeV are used.}
  \begin{ruledtabular}
  \begin{tabular}{cccccccccc}
    Parameter set & $m_{\sigma}$ [MeV] & $m_{\omega}$ [MeV] &
    $g_{\sigma}$ & $g_{\omega}$ & $g_{2}$ [fm$^{-1}$] & $g_{3}$ &
    $c_{3}$ & $g_{\rho}$ \\
    \hline
    NL3-hp ($\Lambda_{\omega}=0$) & $508.194$ & $782.5$ & $10.217$ &
    $12.868$ & $-10.431$ & $-28.885$ & $0$ & $4.461$ \\
    NL3-hp ($\Lambda_{\omega}=0.025$) & $508.194$ & $782.5$ & $10.217$ &
    $12.868$ & $-10.431$ & $-28.885$ & $0$ & $5.376$ \\
    Z271 ($\Lambda_{\omega}=0$) & $505.0$ & $783.0$ & $7.031$ & $8.4065$
    & $-5.4345$ & $-63.691$ & $49.94$ & $4.749$ \\
    Z271 ($\Lambda_{\omega}=0.040$) & $505.0$ & $783.0$ & $7.031$ &
    $8.4065$ & $-5.4345$ & $-63.691$ & $49.94$ & $5.008$ \\
    TM1 ($\Lambda_{\omega}=0$) & $511.198$ & $783.0$ & $10.0289$ &
    $12.6139$
    & $-7.2325$ & $0.6183$ & $71.3075$ & $4.6322$ \\
  \end{tabular}
  \end{ruledtabular}
  \label{tab:param}
\end{table*}

As is well known nowadays, the RMF model Lagrangian with nonlinear
cubic, quartic terms of $\sigma$ bosons, and a quartic one of $\omega$
mesons has achieved remarkable successes in studies on nuclear/hadronic
physics. This type of Lagrangian was initiated
by~\citeauthor{bodmer91:_relat_mean_field_theor_nuclei}%
~\cite{bodmer91:_relat_mean_field_theor_nuclei} and has matured
steadily~\cite{gmuca92:_relat,sugahara94:_relat}.  A
parameter set is determined so as to reproduce the saturation
properties of symmetric nuclear matter and ground-state properties of
typical finite nuclei. A prominent feature of the model is that
scalar and vector self-energies (or Hartree fields) follow the
density-dependence of those obtained by the
Dirac-Brueckner-Hartree-Fock (DBHF) approach even in moderately
supranuclear density. Hence, the model is considered to have the
capability to deal with neutron star matter at such density.
This model Lagrangian is of the form
\begin{equation}
  \label{eq:conv-rmf-lagr}
    \begin{split}
      \mathcal{L} & =
      \bar\psi[i\gamma_\mu\partial^\mu - (M + g_\sigma \sigma)
      - g_\omega \gamma_\mu\omega^\mu 
      - g_\rho \gamma_\mu \bm{\tau}\cdot\bm{b}^\mu ]\psi \\
      & {}+ \frac{1}{2}(\partial_\mu\sigma)(\partial^\mu\sigma)
      {}- \frac{1}{2}m_\sigma^2\sigma^2 - \frac{g_{2}}{3}\sigma^3 -
      \frac{g_{3}}{4}\sigma^4 \\
      & {}- \frac{1}{4}\Omega_{\mu\nu}\Omega^{\mu\nu}
      {}+ \frac{1}{2}m_\omega^2\omega_\mu\omega^\mu +
      \frac{c_{3}}{4}(\omega_\mu \omega^\mu)^2 \\
      & {}- \frac{1}{4}\bm{B}_{\mu\nu} \cdot \bm{B}^{\mu\nu}
      +\frac{1}{2}m_\rho^2 \bm{b}_\mu\cdot\bm{b}^\mu ,
    \end{split}
\end{equation}
where $\Omega_{\mu\nu}=\partial_{\mu}\omega_{\nu} -
\partial_{\nu}\omega_{\mu}$ and
$\bm{B}_{\mu\nu}=\partial_{\mu}\bm{b}_{\nu} -
\partial_{\nu}\bm{b}_{\mu}$. The symbols $\psi$, $\sigma$,
$\omega_{\mu}$, $\bm{b}_{\mu}$, $M$, $m_{\sigma}$, $m_{\omega}$, and
$m_{\rho}$ signify the fields of nucleons, $\sigma$ bosons, $\omega$
mesons, $\rho$ mesons, the masses of nucleons, $\sigma$ bosons,
$\omega$ mesons, and $\rho$ mesons, respectively. We call
Eq.~\eqref{eq:conv-rmf-lagr} ``standard RMF Lagrangian.''


As an extension of the well-established standard RMF model, the
effective field theory (EFT)~\cite{serot97:_recen} provides a modern
aspect of the RMF model; an energy density functional obtained from
the RMF Lagrangian approximates the \emph{exact} energy density
functional of the ground state of a hadronic system in the sense of
the density functional theory (DFT). Adding apt interaction terms,
absent in the standard one, to the RMF functional advances it to the
exact functional little by little.
In this EFT-inspired approach, mesons and baryons are not taken as
elementary degrees of freedom, so that a Lagrangian composed of
corresponding fields is allowed to be nonrenormalizable. This leads
to, in principle, unrestricted inclusion of \emph{any} terms of meson
self-interactions consistent with symmetries of the underlying theory,
QCD. These terms are completely absent in the standard RMF Lagrangians
that respect renormalizability.
Recently, \citeauthor{horowitz01:_neutr_star_struc_neutr_radius}
proposed an RMF Lagrangian along with concepts and methods of the EFT
and the DFT.
They extended the standard RMF Lagrangian with a nonlinear
$\omega$-$\rho$ coupling particularly from the viewpoint of the
symmetry energy. The new coupling paves the way to change the density
dependence of the symmetry energy, and hence the proton fraction in
neutron star matter.
For the extended Lagrangian, the present study employs an
additional interaction Lagrangian of the
form~\cite{horowitz01:_neutr_star_struc_neutr_radius}
\begin{equation}
  \label{eq:eft-insp-lagr}
  \mathcal{L}_\mathrm{EFT} = \mathcal{L}
  + 4 g_{\rho}^{2} \bm{b}_{\mu} \cdot \bm{b}^{\mu}
   \Lambda_{\omega} g_{\omega}^{2} \omega_{\nu} \omega^{\nu},
\end{equation}
where $\Lambda_{\omega}$ signifies a nonlinear coupling
constant. We name Eq.~\eqref{eq:eft-insp-lagr} ``EFT-inspired
Lagrangian.'' The important thing is that the Lagrangian
\eqref{eq:eft-insp-lagr} gives the symmetry energy coefficient
\begin{equation}
  \label{eq:sym-energy}
  a_\textrm{sym} =
  \frac{k_\mathrm{F}^{2}}{6\sqrt{k_\mathrm{F}^{2} + {M^{\ast}}^{2}}}
  + \frac{g_{\rho}^{2}}{3\pi^{2}}
  \frac{k_\mathrm{F}^{3}}{m_{\rho}^{\ast 2}}.
\end{equation}
The symbol $m_{\rho}^{\ast}$ denotes the effective mass of $\rho$
mesons
\begin{equation}
  \label{eq:eff-rho-mass}
  m_{\rho}^{\ast 2} = m_{\rho}^{2}
  + 8 g_{\rho}^{2} 
  \Lambda_{\omega} g_{\omega}^{2} {\langle \omega_{0} \rangle}^{2},
\end{equation}
where $\langle \mathcal{O} \rangle$ represents an expectation value of
a field $\mathcal{O}$ for the ground state of the system.
In this study, we pick out the \mbox{NL3-hp} and Z271 parameter
sets~\cite{horowitz01:_neutr_star_struc_neutr_radius,%
carriere03:_low_mass_neutr_stars_equat}; each set gives the lower or
upper limit of commonly accepted range of the effective nucleon mass
in symmetric nuclear matter at $\rho_{0}$, $M^{\ast} \simeq$
(0.6--0.8)$M$. Most RMF parameter sets yield the mass within this
range. Note that we rename the NL3 set
by~\citeauthor*{horowitz01:_neutr_star_struc_neutr_radius} in
Ref.~\cite{horowitz01:_neutr_star_struc_neutr_radius} here because of
their slight modification of the well-known NL3 parameter
set~\cite{lalazissis97:_new_lagran}.
We will also mention the TM1 parameter set~\cite{sugahara94:_relat}
for the standard RMF Lagrangian in addition to the above two sets.
Table~\ref{tab:param} shows the parameter sets.


In neutron star matter composed by the models, we solve the gap
equation for the $^{1}S_{0}$ proton pairing correlation at zero
temperature using the Bonn-\textit{B} potential~\cite{machleidt89:_meson} as a
particle-particle (\mbox{\textit{p-p}}) channel interaction $\bar{v}$,
\begin{eqnarray}
  \label{eq:gap-eq}
  \Delta (p) & = & \displaystyle {}- \frac{1}{8 \pi^{2}}
  \int_{0}^{\infty}
  \frac{\Delta (k)}{\sqrt{(E_{k} - E_{k_\mathrm{F}})^{2} 
      + \Delta^{2} (k)}} \nonumber \\
  & & \qquad \qquad \quad \times \;
  \bar{v}(p, k) \: k^{2} dk, 
\end{eqnarray}
where $E_{k} = \sqrt{k^{2} + {M^{\ast}}^{2}} + g_{\omega} \langle
\omega_{0} \rangle + g_{\rho} \langle b_{0}^{(3)} \rangle$. The
\mbox{\textit{p-p}} channel interaction between protons $\bar{v}(p,k)$
is obtained through angular integration with respect to the angle
between linear momenta $\mathbf{p}$ and $\mathbf{k}$. This serves as a
process of projecting out the \textit{S}-wave component of the
interaction. Its original form is nothing but the antisymmetrized
matrix element of the employed interaction $V$, which is defined by
\begin{equation}
\label{eq:antisym-vpp}
\bar{v}(\mathbf{p},\mathbf{k})
=\langle \mathbf{p}s',\widetilde{\mathbf{p}s'}\vert V\vert
           \mathbf{k}s,\widetilde{\mathbf{k}s}\rangle
   -\langle \mathbf{p}s',\widetilde{\mathbf{p}s'}\vert V\vert
           \widetilde{\mathbf{k}s},\mathbf{k}s\rangle ,
\end{equation}
where a tilde over its argument denotes time reversal.
We use the lowest approximation as to the \mbox{\textit{p-p}}
channel interaction to study exclusively the effect of the bulk
properties on superfluidity. It is, however, known that the
polarization effects bring about significant reduction of the gap
within nonrelativistic frameworks~%
\cite{ainsworth89:_effec_inter_energ_gaps_low,schulze96:_medium}.


The models employed here should be regarded as hybrid models because
we use different interactions in the particle-hole
(\mbox{\textit{p-h}}) and the particle-particle channel. This prevents
ambiguity in choosing the pairing interaction. Thus we concentrate on
the study of the effect of the Dirac effective mass and the proton
fraction on the proton superfluidity.

\section{Results and discussion}
\label{sec:results-discussion}


\begin{figure}[tbp]
  \centering
  \includegraphics[width=7.5cm,keepaspectratio]{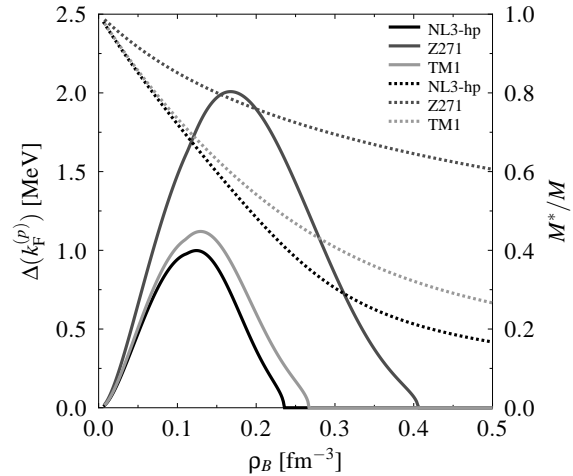}
  \caption{$^{1}S_{0}$ proton pairing gaps (left scale, solid curves)
    and Dirac effective masses (right scale, dashed curves) as
    functions of baryon density in neutron star matter using the
    standard RHB model with the \mbox{NL3-hp},
    Z271~\cite{horowitz01:_neutr_star_struc_neutr_radius}, and
    TM1~\cite{sugahara94:_relat} parameter sets.}
  \label{fig:gap+efm}
\end{figure}

To begin with, we calculate the $^{1}S_{0}$ pairing gaps and the Dirac
effective masses of protons \emph{without} the isovector nonlinear
coupling. The results are shown in Fig.~\ref{fig:gap+efm}, which
apparently shows that the gaps strongly depend on the Dirac effective
masses. The smaller the mass is, the smaller the pairing gap is.
The \mbox{NL3-hp} set makes symmetric nuclear matter saturated with the Dirac
effective mass of $0.59M$, while the Z271 set does with that of $0.80M$.
According to Ref.~\cite{carriere03:_low_mass_neutr_stars_equat}, both
sets give similar properties of typical finite nuclei and low-mass
neutron stars on the one hand, different properties of typical neutron
stars of mass $1.4M_{\odot}$ follow from the sets on the other hand.
For comparison, we also present the result of the TM1 set in
Fig.~\ref{fig:gap+efm} with the light gray curves. The effective mass
obtained by the TM1 set lies in between those by the other two sets, so
does the pairing gap; it is consistent with the above statement as
well.
In view of the pairing correlation in neutron star matter, using the
different \mbox{\textit{p-h}} interactions and the same
\mbox{\textit{p-p}} interaction gives the significant difference shown
in Fig.~\ref{fig:gap+efm}. Hence, this suggests that we can use the
strong correlation as a guide to narrow down the parameter sets
favorable for a calculation such as the present study.

\begin{figure}[tbp]
  \centering
  \includegraphics[width=6.8cm,keepaspectratio]{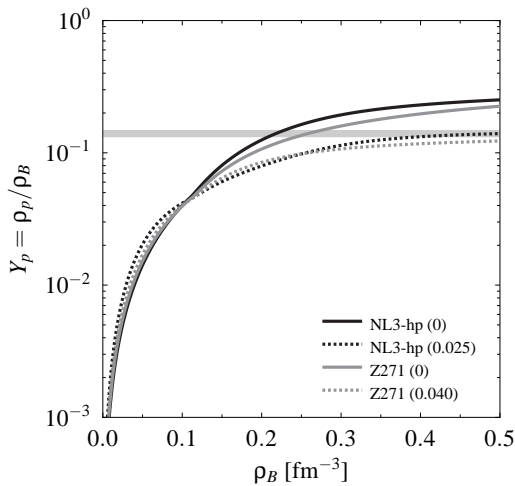} 
  \caption{Proton fractions of the four distinct models of neutron
    star matter as functions of baryon density. The thin region
    hatched in gray represents a threshold of the proton fraction for
    the direct URCA process, about 13--15~\%. The number in parenthesis
    in the legend indicates the value of $\Lambda_{\omega}$.}
  \label{fig:part-frac}
\end{figure}

Next, we show the proton fractions $Y_{p} = \rho_{p}/\rho_{B}$ in
Fig.~\ref{fig:part-frac} for the variations of Lagrangian. The solid
black curve denotes the proton fraction obtained by the \mbox{NL3-hp}
set, the solid gray curve by the Z271 set, both \emph{without} the new
nonlinear coupling. Their characteristic is the large proton fraction,
which is the typical result obtained from the standard RMF Lagrangian.
The result obtained by the TM1 set is similar to the one by the
\mbox{NL3-hp} set without the coupling (and hence is omitted hereafter).
In contrast, using finite $\Lambda_{\omega}$, namely, the EFT-inspired
RMF Lagrangian, we have obtained smaller proton fractions at densities
corresponding to the core region of neutron stars, $\rho_{B} \gtrsim
0.1$ fm$^{-3}$, than using $\Lambda_{\omega}=0$; these fractions are
drawn with dashed curves in Fig.~\ref{fig:part-frac}. As the original
intent of this kind of
Lagrangian~\cite{horowitz01:_neutr_star_struc_neutr_radius}, it yields
gentle increase of the fraction at these relevant densities owing to
the symmetry energy adjustable via the isovector nonlinear coupling
$\Lambda_{\omega}$ indicated by Eqs.~\eqref{eq:sym-energy}
and~\eqref{eq:eff-rho-mass}.
Not shown are curves of the proton fraction that lie in between the two
extreme conditions $\Lambda_{\omega}=0$ and $0.025$ ($0.040$) for the
\mbox{NL3-hp} (Z271) set, when it is varied between them. To wind up
the account of Fig.~\ref{fig:part-frac}, we should note how each curve
of the proton fraction depends on baryon density, with its influence
on the pairing correlation in mind.

\begin{figure}[tbp]
  \centering
  \includegraphics[width=7cm,keepaspectratio]{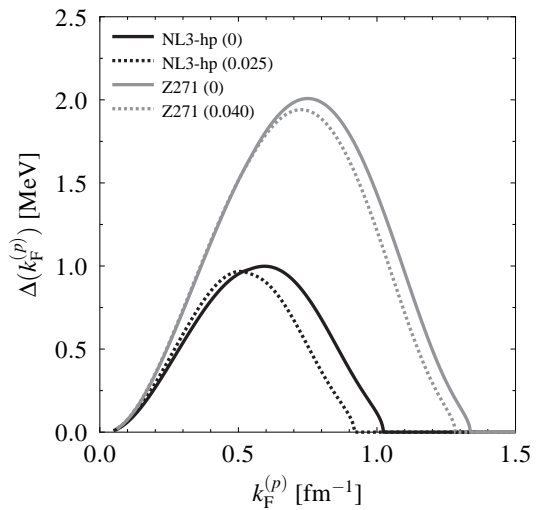}
  \caption{$^{1}S_{0}$ proton pairing gaps as functions of the proton
    Fermi momentum in neutron star matter using the RHB models with
    the \mbox{NL3-hp} and Z271 parameter sets with and without
    $\Lambda_{\omega}$. The legend is the same as in
    Fig.~\ref{fig:part-frac}.}
  \label{fig:kfp-gap}
\end{figure}

To offer more direct information on the $^{1}S_{0}$ proton pairing
gaps at the proton Fermi surface and their relation to the proton
fractions, we present the gaps as functions of the proton Fermi
momentum in Fig.~\ref{fig:kfp-gap}.
An apparent difference between the gaps obtained with the
\mbox{NL3-hp} and Z271 sets stems, again, from the difference of the
Dirac effective masses between the sets, as shown in
Fig.~\ref{fig:gap+efm}. In addition, the two curves for each parameter set
with and without $\Lambda_{\omega}$ reflect the increase rate of the
proton fraction as functions of baryon density (see
Fig.~\ref{fig:part-frac}).  Comparing the property of protons at the
same Fermi momentum explains further details about that:
With $\Lambda_{\omega}$, where the rate is moderate at the relevant
densities, the proton Cooper pairs are immersed in denser background
than those without $\Lambda_{\omega}$, where the rate is
rapid. Meanwhile, the higher the baryon density is, the smaller the
Dirac effective mass of protons is. We thus obtain the smaller pairing
gaps at the same Fermi surface for the parameter sets with
$\Lambda_{\omega}$, namely, for the EFT-inspired Lagrangian.

\begin{figure}[tbp]
  \centering
  \includegraphics[width=7cm,keepaspectratio]{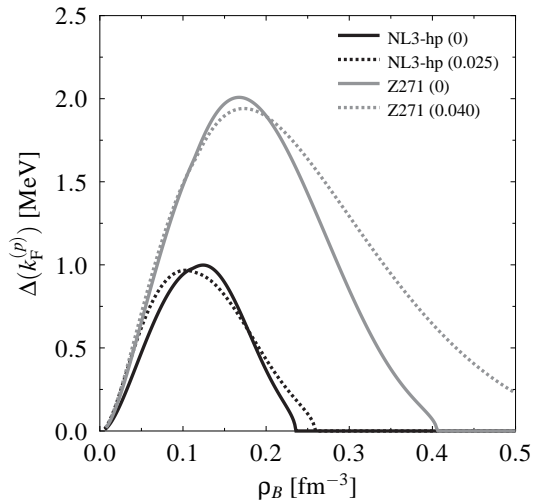} 
  \caption{$^{1}S_{0}$ proton pairing gaps as functions of baryon
    density in neutron star matter using the RHB models with the
    \mbox{NL3-hp} and Z271 parameter sets with and without
    $\Lambda_{\omega}$. Solid curves and the legend are the same as in
    Figs.~\ref{fig:gap+efm} and~\ref{fig:part-frac}, respectively.}
  \label{fig:gap-nliv}
\end{figure}

We compare the pairing gaps as functions of baryon density obtained
with and without the isovector nonlinear coupling $\Lambda_{\omega}$
in Fig.~\ref{fig:gap-nliv}. Since the density dependence of the Dirac
effective masses obtained from the respective parameter sets is almost
the same for any values of $\Lambda_{\omega}$, the obtained gaps
reflect the respective proton fractions presented in
Fig.~\ref{fig:part-frac}.  As a whole, the pairing gaps survive in
higher density region for the EFT-inspired Lagrangian with finite
$\Lambda_{\omega}$, and have lower peaks than for the standard
Lagrangian with $\Lambda_{\omega}=0$. Figures~\ref{fig:part-frac}
and~\ref{fig:gap-nliv} give indications as follows.%

(1) Up to $\rho_{B} \simeq 0.05$~fm$^{-3}$, the proton fractions are
almost the same except the one of the \mbox{NL3-hp} set with finite
$\Lambda_{\omega}$, which is a little larger than the others
(see Fig.~\ref{fig:part-frac}).  The enhancement of the gap
due to the large proton fraction cancels out the smallness of the
effective mass given by the \mbox{NL3-hp} set. Hence, the pairing gap
obtained by the \mbox{NL3-hp} set with $\Lambda_{\omega}=0.025$ is
almost the same as those by the Z271 set.  We note, however, that this
region roughly corresponds to the inner crust of neutron stars, where
the $^{1}S_{0}$ proton pairing is much weaker than the neutron
counterpart.

(2) At $\rho_{B} \approx 0.1$~fm$^{-3}$, as the proton fractions of
the Z271 set start to deviate from each other, so do the pairing gaps
obtained by the set. As to the \mbox{NL3-hp} set, the proton fractions
for $\Lambda_{\omega}=0$ and $0.025$ nearly coincide there, so do the
pairing gaps.

(3) For $0.1$~fm$^{-3} \lesssim \rho_{B} \lesssim 0.2$~fm$^{-3}$, the
high proton fractions make the corresponding pairing gaps large. This
implies that richness of protons favors the pairing correlation by
taking advantage of an attractive part of the proton-proton
interaction in this region.

(4) At $\rho_{B} \approx 0.2$~fm$^{-3}$, an opposite situation to the
previous case (3) is realized; namely, the high proton fractions make
the corresponding pairing gaps small. This implies that richness of
protons disfavors the pairing correlation by virtue of a repulsive
part of the proton-proton interaction in this region. As a result of
the difference between the effective masses, the pairing gap given by
the Z271 set is still very large.
To sum up the discussion of Fig.~\ref{fig:gap-nliv}, the more gently
protons multiply in the core region of neutron stars, up to the higher
density superfluidity survives and the somewhat smaller the peak of
the gap one obtains.
\begin{figure}[tbp]
  \centering
  \includegraphics[width=7cm,keepaspectratio]{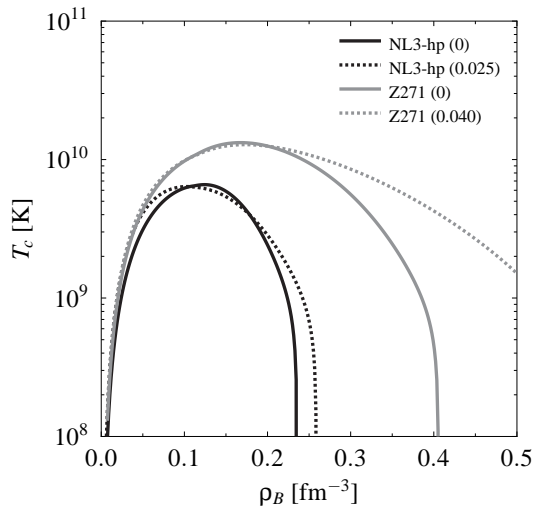} 
  \caption{Critical temperatures of $^{1}S_{0}$ proton superfluids as
    functions of baryon density in neutron star matter using the RHB
    models with the \mbox{NL3-hp} and Z271 parameter sets with and
    without $\Lambda_{\omega}$. The legend is the
    same as in Fig.~\ref{fig:part-frac}.}
  \label{fig:gap-Tc}
\end{figure}
\begin{figure}[tbp]
  \centering
  \includegraphics[width=7cm,keepaspectratio]{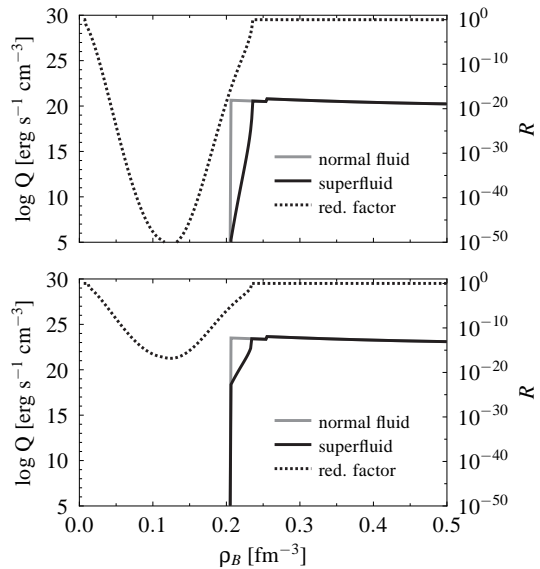}
  \caption{Neutrino emissivities of the nucleon direct URCA process
    (left scale, solid curves) and the reduction factor for
    $^{1}S_{0}$ proton pairing (right scale, dashed curve) as
    functions of baryon density in neutron star matter for the NL3-hp
    parameter set without $\Lambda_{\omega}$. Solid gray curves
    represent the neutrino emissivities in normal fluid. The upper
    panel is drawn for the internal temperature $T = 1.0 \times
    10^{8}$~K, the lower for $T = 3.0 \times 10^{8}$~K.}
  \label{fig:nuemit-NL3hp}
\end{figure}
\begin{figure}[tbp]
  \centering
  \includegraphics[width=7cm,keepaspectratio]{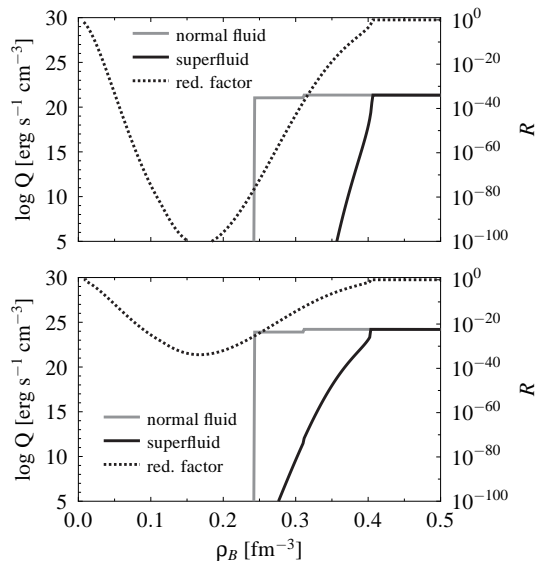}
  \caption{Same as Fig.~\ref{fig:nuemit-NL3hp}, but for the Z271
    parameter set without $\Lambda_{\omega}$. Also note that the right
    scales are different from those of Fig.~\ref{fig:nuemit-NL3hp}.}
  \label{fig:nuemit-Z271}
\end{figure}

Let us here present three figures relevant to the physics of neutron
stars. One is Fig.~\ref{fig:gap-Tc}, which represents the density
dependence of the critical temperatures of the superfluids.  The
critical temperatures $T_{c}$ are obtained by the universal relation for a
weak-coupling BCS superconductor at $T=0$~K, $k_\mathrm{B} T_{c} =
0.57 \Delta(k_\mathrm{F}^{(p)};~T=0)$, where $k_\mathrm{B}$ is the
Boltzmann constant. Since the temperature inside evolved neutron
stars is about $10^{8}$~K, Fig.~\ref{fig:gap-Tc} signifies that the
superfluidity is likely to exist in the all cases considered.
The other two are Figs.~\ref{fig:nuemit-NL3hp}
and~\ref{fig:nuemit-Z271}, which depict the neutrino emissivities $Q$
and their reduction factor $R$ by the $^{1}S_{0}$ proton superfluidity
as functions of baryon density, drawn at the internal temperatures $T
= 1.0 \times 10^{8}$~K and $3.0 \times 10^{8}$~K.
We now take, for example, the nucleon direct URCA process as a cooling
agent of neutron stars.
The neutrino emissivity of the nucleon direct URCA process in normal
fluid $Q_{0}$ has the following form in units of erg s$^{-1}$
cm$^{-3}$~\cite{lattimer91:_direc_urca_proces_neutr_stars}:
\begin{equation}
  \label{eq:emissivity}
  Q_{0} = 7.55 \times 10^{30} \mu_{e} T_{9}^{6}
  \frac{{M_{n}^{\ast}}^{2} {M_{p}^{\ast}}^{2}}{M^{2}}
  \theta(p_{\mathrm{F} e} + p_{\mathrm{F} p} - p_{\mathrm{F} n}),
\end{equation}
where $\mu_{e}$ is the chemical potential of electrons (and muons due
to the condition of beta equilibrium), $T_{9}$ is the temperature in
units of $10^{9}$~K, and $\theta(p_{\mathrm{F} e} + p_{\mathrm{F} p} -
p_{\mathrm{F} n})$ is the triangle condition for particle momenta.
After the appearance of muons, they also contribute to the neutrino
emission so that the emissivity $Q_{0}$ just doubles, which manifests
itself as the small kinks after the ignition of the process involving
electrons (abrupt increase of the emissivities) in
Figs.~\ref{fig:nuemit-NL3hp} and~\ref{fig:nuemit-Z271}.
It should be noted that we approximate the effective masses
$M_{n}^{\ast} = M_{p}^{\ast} = M^{\ast}$ in Eq.~\eqref{eq:emissivity}:
To distinguish between them in the RMF model, we must introduce
isovector-scalar mesons and shortly discuss this issue later.
The neutrino emissivity in superfluid is thus written as $Q = Q_{0}
R$, the reduction factor used here being
\begin{eqnarray}
  \label{eq:sup-factor}
  R & \simeq & \exp \left[ - \Delta(k_\mathrm{F}^{(p)}; T) /
    k_\mathrm{B} T \right] \nonumber \\
  & \simeq & \exp \left[ - \Delta(k_\mathrm{F}^{(p)}; T=0) /
    k_\mathrm{B} T \right] \nonumber \\
  & \simeq & \exp \left[ - 1.76 T_{c}(k_\mathrm{F}^{(p)}) / T \right],
\end{eqnarray}
which is the form often used in the past studies.  It should be
mentioned that the reduction factor of the form
\eqref{eq:sup-factor} leads to an overestimate of the suppressive
effect~\cite{yakovlev01:_neutr,takatsuka04:_baryon_super_neutr_emiss_neutr_stars}.
Although we have employed the pairing gaps at $T=0$~K for a
qualitative estimate in the present study, one should use the pairing
gaps at finite temperatures for a quantitative study of thermal
evolution of neutron stars.

As shown in Figs.~\ref{fig:gap+efm}~and~\ref{fig:part-frac}, both the
NL3-hp and the Z271 set with $\Lambda_{\omega}=0$ yield the relatively
large proton fraction. Meanwhile, the pairing gap for the NL3-hp set
closes around the density at which the direct URCA process takes
effect. Therefore, the reduction effect is not much large, which is
clear for $T = 3.0 \times 10^{8}$~K as shown in
Fig.~\ref{fig:nuemit-NL3hp}.
By comparison, the superfluid range for the Z271 set is wide
enough to cover the density region where the direct URCA process is
turned on. Hence the process is strongly suppressed by the interior
proton superfluid as shown in Fig.~\ref{fig:nuemit-Z271}.
For both sets with finite $\Lambda_{\omega}$, on the other hand, the
density regions of the superfluidity and the direct URCA process do
not overlap each other; the latter starts at $\rho_{B} = 0.442$
fm$^{-3}$ with $Y_{p} = 0.136$ for the NL3-hp set and at $\rho_{B} =
1.028$ fm$^{-3}$ with $Y_{p} = 0.141$ for the Z271 set. Thus the
superfluidity is of no suppressive effect on the direct URCA process
for both with finite $\Lambda_{\omega}$.



Now we develop a brief discussion on the cooling of neutron stars
relative to the above results.
From the aspect of surface temperature of neutron stars, we can
classify observed neutron stars into hotter and colder ones. Broadly
speaking, this means that there exist two major cooling scenarios; the
so-called ``standard'' and ``nonstandard''
cooling~\cite{tsuruta98:_therm_proper_and_detec_of_neutr_stars}. The
standard cooling is dominated by the modified URCA process, which
cools neutron stars slowly and thus results in hotter stars. In
contrast, the nonstandard cooling includes the nucleon/hyperon direct
URCA processes, via which neutron stars cool faster and lead to colder
stars, in conjunction with the suppression due to
superfluidity. Concerning this classification, the pulsar PSR
J0205+6449 was recently discovered in the supernova remnant 3C~58
by~\citeauthor{murray02:_discov_of_x_ray_pulsat}%
~\cite{murray02:_discov_of_x_ray_pulsat}.
\citeauthor{slane02:_new_const_on_neutr_star} subsequently deduced a
strong upper limit on the surface
temperature~\cite{slane02:_new_const_on_neutr_star}. They showed that
the limit falls below the prediction of the standard cooling model,
which suggests that PSR J0205+6449 seems to follow the nonstandard
cooling scenario.
In the first place, the proton fraction higher than about 13--15~\%
(shown by the hatched region in Fig.~\ref{fig:part-frac}) activates
the direct URCA process involving nucleons in neutron star matter with
the standard RMF Lagrangian. However, our results from the Z271 set
show that the proton pairing gap is large enough to suppress the
process, so that the resulting cooling of neutron stars is not much
effective to meet observational data of colder neutron stars. This
calls for the cooling that results from hyperons or meson condensates;
the latter consequently means coexistence of superfluidity and meson
condensations.
In the second place, using the EFT-inspired RMF Lagrangian with the
Z271 set, neutron star matter is composed with the proton fraction
lower than the direct URCA threshold up to $\sim 7\rho_{0}$. This
prohibits neutron stars from cooling by the direct URCA process
involving nucleons; the modified URCA process dominates their cooling
instead, given the present constituent particles. In this regard,
\citeauthor*{horowitz02:_const_urca} studied the direct URCA process
in detail using the EFT-inspired RMF Lagrangian with intermediate
values of $\Lambda_{\omega}$~\cite{horowitz02:_const_urca}.
Bear in mind that we have put aside the Cooper pair breaking and
formation processes%
~\cite{flowers76:_neutr_pair_emiss_from_finit,voskresensky87:_descr_keldy}
that work as a cooling accelerator, while we have taken superfluidity
into account as a cooling retardant alone: Since the neutrino
emissivity of these processes sensitively depends on the density and
the temperature dependence of the pairing gap, our approximate
treatment at the moment does not allow us to estimate this emissivity
and to determine the scenario that the pulsar follows.

Finally, we should note preceding works on the $^{1}S_{0}$ proton
superfluidity in neutron star matter. According to them,
nonrelativistic approaches and the Dirac-Brueckner-Hartree-Fock
approach give similar results with the maximal pairing gap
$\Delta_\mathrm{max} \lesssim 1.0$~MeV and the closing density
$\rho_{c} \sim 0.4$~fm$^{-3}$%
~\cite{takatsuka97:_nucleon_super_neutr_star_core,%
elgaroey96:_model_s_bonn,elgaroey96:_super}. The discrepancy between
them and our results having been exposed, with the reservation of the
lowest approximation, we now need information to narrow down further
RMF parameter sets for the purpose like the present study. In this
context, extracting the EOS from direct observation of the stars and
experiments of neutron-rich nuclei is indispensable.

\section{Summary and conclusions}
\label{sec:summary}

We have investigated the $^{1}S_{0}$ proton pairing in neutron star
matter using the relativistic Hartree-Bogoliubov model.
Since proton Cooper pairs reside in extremely dense surroundings
inside neutron stars, special care should be taken of the effects
of the bulk properties, as well as pairing interactions and levels
of the approximation. Therefore, we have studied the effects on the
pairing correlation using the lowest approximation with respect to the
pairing interaction.
We summarize and conclude the following: First, using the standard RMF
parameter, we have obtained that the maximal pairing gap is about
1--2~MeV, dependent on the parameter sets used, which is comparable or
larger than the values obtained in the preceding studies. This clearly
shows the strong correlation between the effective mass of nucleons
and the pairing gap, which was also concluded within nonrelativistic
models.  Although the relativistic model tends to have a smaller
effective mass of nucleons (the Dirac effective mass) than those in
nonrelativistic models, the value of the maximal gap is not much small
against general expectations.
Second, we have studied the problem also using the extended Lagrangian
proposed in line with a concept of the EFT
by~\citeauthor{horowitz01:_neutr_star_struc_neutr_radius}%
~\cite{horowitz01:_neutr_star_struc_neutr_radius}. The proton fraction
obtained can be controlled by the additionally introduced parameter
$\Lambda_{\omega}$. We have found that the more slowly protons
increase with density in the core region of neutron stars, the wider
the superfluid range and the slightly lower the peak of the gap become.

In this study, we have ruled out direct effects of the bulk properties
on the \mbox{\textit{p-p}} channel interaction: For instance,
self-consistent (or in-medium) Dirac spinors in the
\mbox{\textit{p-p}} channel are not used here though they probably
affect properties of the pairing correlation at high
density~\cite{tanigawa03:_possib_lambd_hartr_bogol}.
Furthermore, we have included no isovector-scalar meson, which may
play important roles in high-density and isovector physics through
larger effective mass of protons than that of
neutrons~\cite{kubis97:_nuclear}. In general, the isovector-scalar
meson increases protons in the high-density region, which is an effect
opposite to that of the EFT-inspired Lagrangian employed here. Last
but not least, having clarified the effects of the bulk properties on
superfluidity by solving the gap equation at zero temperature, we are
prepared for solving it at finite temperatures. It will be mandatory to
consider these issues.
With attention to the bulk properties of neutron star matter, a study
of baryon superfluidity therein and accompanying neutrino emissivities
is in progress.

\begin{acknowledgments}
  One of us (T.T.) is grateful to the Japan Society for the Promotion
  of Science for research support and the members of the research
  group for manybody theory of hadron systems at the Japan Atomic
  Energy Research Institute (JAERI) for fruitful discussions.
\end{acknowledgments}

\end{document}